\documentclass[runningheads]{llncs}
\usepackage[utf8]{inputenc}
\usepackage[T1]{fontenc}
\usepackage{amsmath}
\usepackage{amsfonts}
\usepackage{amssymb}
\usepackage{graphicx}
\usepackage{hyperref}
\usepackage{booktabs}
\usepackage{algorithm}
\usepackage{algorithmic}
\usepackage{geometry}
\usepackage{url}
\usepackage{orcidlink}

\begin{document}

\title{How Worrying Are Privacy Attacks Against
Machine Learning?} 

\titlerunning{How Worrying Are Privacy Attacks Against ML?}

\author{Josep Domingo-Ferrer\inst{1,2}\orcidlink{0000-0001-7213-4962}}

\authorrunning{Josep Domingo-Ferrer}

\institute{Universitat Rovira i Virgili,\\ Department of Computer Engineering and Mathematics,\\
        CYBERCAT-Center for Cybersecurity Research of Catalonia,\\ Av. Pa\"{\i}sos Catalans 26, 43007 Tarragona, Catalonia\\
\email{josep.domingo@urv.cat} \and
LAAS-CNRS, Universit\'e de Toulouse\\ 
7 Av. du Colonel Roche, 31400 Toulouse, France}

\maketitle

\begin{abstract}
In several jurisdictions,
the regulatory framework on the release and sharing of personal data is
being extended to machine learning (ML). The implicit assumption is
that disclosing a trained ML model entails a privacy
risk for any personal data used in training comparable to directly releasing those data. However, given a trained model, it is necessary to mount a {\em privacy attack} to make inferences on the training data. In this concept paper, we examine the main families of privacy attacks against predictive and generative ML, including membership inference attacks (MIAs), property inference attacks, and reconstruction attacks. Our discussion shows that most of these attacks seem less effective in the real world than what a {\em prima facie} interpretation of the related literature could suggest. 
\end{abstract}

\section{Introduction}
\label{intro}

The main regulations in the EU that affect the development of AI are
the General Data Protection Regulation (GDPR) and the EU Artificial
Intelligence Act. Both were conceived before the boom of generative AI in
2022. Furthermore, the EU has announced the implementation of a
Code of Practice for general purpose AI models~\cite{codepractice}. 
Outside Europe, in 2023 President Biden had signed Executive Order 14110, which committed the USA to a strong regulation of the development and use of AI along similar lines as the European regulation. However, in 2025 President Trump has signed Executive Order 14179, which basically revokes Biden’s order and removes all AI regulations, allegedly to ``remove barriers to American leadership in artificial intelligence''. 

Given the above situation and the fact that Chinese regulations on AI are more
focused on protecting the government than the citizens from AI, the EU remains the world's only major economic bloc committed to trustworthy AI. At the same time, the EU lags behind the USA and China from the AI technology point of view.

For the European AI industry to be able to catch up with its competitors in spite of a more strict regulatory framework, it is extremely important to make sure that regulations are not more strict than required to preserve the values of trustworthy AI, and in particular privacy. Unfortunately, this does not seem to be the case today. The EU regulations assume that {\em any} disclosure
might cause a breach of privacy. 
In particular, there is an implicit assumption that the disclosure of a trained machine learning (ML) model entails a privacy risk for any personal data used in training comparable to the direct release of those data.

This overcautious approach is probably
due to the rushed inclusion of generative AI in the legal texts, and it may
lead to adopting countermeasures that increase the training overhead
and decrease the accuracy of models. For example, differential
privacy~\cite{dwork2006calibrating} is a commonly proposed countermeasure that
can cause two-digit drops in model accuracy if applied with
meaningful privacy parameters. This seriously compromises the performance
and competitiveness of the models and might be {\em unnecessary}
if risks can be demonstrated to be overestimated.

\subsection*{Contribution and plan of this article}

There is a fundamental privacy difference between releasing an ML model trained on personal data and directly releasing those training data. 
If only the trained ML model is disclosed, it is necessary to mount a {\em privacy attack} to make any inferences on the training data. 
In this paper, we discuss how effective the privacy attacks
proposed in the literature against predictive and generative ML are in 
{\em real-world} conditions.
Specifically, we cover membership inference attacks (MIAs), property
inference attacks, and reconstruction attacks.

Our assessment concentrates on the disclosure potential
of those attacks at the conceptual level, rather than
on the analysis of the internals of the various attack techniques.
We aim to uncover fundamental limitations of privacy attacks.

Section~\ref{background} gives a background on privacy
disclosure. Section~\ref{MIA} is devoted to membership inference
attacks.
Section~\ref{property} discusses property inference attacks.
Section~\ref{reconstruction} deals with reconstruction attacks.
Conclusions are drawn in Section~\ref{conclusion}.

\section{Background on privacy disclosure}
\label{background}

For many years, the literature on database privacy~\cite{sdc-book} 
has used the notion
of disclosure risk, in order to measure to what extent the release
of data sets and statistical output puts sensitive information at risk
of being disclosed. This notion remains relevant in the machine
learning domain.

Two types of disclosure have usually been considered~\cite{sdc-book}:
\begin{itemize}
\item {\em Identity disclosure} means that the attacker
is able to link some
unidentified piece of data released with the subject (individual) to whom it corresponds. This linkage is also called {\em re-identification}.
\item {\em Attribute disclosure} means that the attacker can
determine the value of a confidential attribute ({\em e.g.},
income, diagnosis, etc.) for a target subject
with great precision {\em after} seeing the released data.
\end{itemize}

In tabular data, reidentification occurs trivially if the released data
contain {\em personal identifiers} (such as passport numbers). That is
why identifiers should never be released. However,
re-identification is also possible by {\em quasi-identifiers}
(for example, gender, job, zipcode, age) that do not uniquely
identify the subject, but whose combination may because
they may be present in public identified databases such as electoral
rolls. Finally, {\em confidential attributes}
(income, diagnosis, etc.) reveal sensitive information
about subjects when they can be unequivocally linked to them. 

Identification and attribute disclosure can occur independently. 
A record can be reidentified but, if it contains no confidential
attribute, no attribute disclosure occurs. Similarly, if the attacker
can only determine a set of $k > 1$ records that might correspond to the
target subject, but there is a confidential attribute whose values
over those $k$ records are very similar, then attribute disclosure
has occurred without reidentification. 

{\em Membership disclosure} has been proposed as a third type
of disclosure in machine learning~\cite{shokri2017membership}.
Its purpose is to determine whether a given data point was
included in the data set used to train a certain ML model.
Thus, in a membership inference attack (MIA), the attacker
does not try to discover to whom the point corresponds
(which would be reidentification) or
to find the value of any confidential
attribute about the subject to whom the point corresponds
(which would be attribute disclosure).

Thus, it can be argued that membership disclosure is weaker
than identity or attribute disclosure. However, 
{\em if all subjects included in the training data set
are known to share a sensitive
condition, attribute disclosure can result from membership disclosure.} 
For example, if all subjects whose data are used for
training suffer from a certain disease, then discovering membership
for a target subject leads to attribute disclosure: the target
suffers from that disease. Note that this is true even
if there was no explicit attribute `Disease' in the training data set.

\section{Membership inference attacks}
\label{MIA}

MIAs are the most common attack employed to assess the privacy
of training data in machine learning.
Rather than analyzing the operation of specific MIAs proposed
in the literature, in this section we will focus on the disclosure potential of a generic MIA depending on the data used to train the model
under attack. 

Let us introduce a running example. Assume that an attacker Alice
wants to perform an MIA on an ML model to determine whether
the attacker's neighbor Neil was a member of the data used to train
the model. 

Two properties of the training data are simultaneously
required for the MIA to allow unequivocal inferences:
\begin{itemize}
\item {\em Exhaustivity.} 
Unless the training data were an exhaustive sample of a population
(which is very rare), membership inferences cannot be unequivocal.
In other words, the membership revealed by an MIA to a non-exhaustive training set could be plausibly denied.
In the running example, if Alice finds that one or several records containing the same quasi-identifier values known to her about Neil were members of the training data, she cannot be absolutely sure that Neil was a member. The reason is that perhaps Neil was not included in the training data, and the putative members she found are just people sharing Neil's quasi-identifier values. Hence, Neil could plausibly deny being a member. 
On the other hand, if the training data set is exhaustive, membership 
is trivial and no MIA is really needed: every existing record (and Neil's
in particular) is a member.
\item {\em Non-diversity of unknown attributes.} Since inferring
membership to an exhaustive sample is not a real discovery, let us examine whether at least it can bring attribute disclosure. If there are several member records matching
the attribute values known to the attacker, and the unknown attributes
among those records differ significantly, then no attribute disclosure occurs.
In the running example, if Alice finds that two or more records containing 
the same quasi-identifiers known to her about Neil were members of the 
(exhaustive) training data, but the confidential attribute Income unknown to Alice take clearly different values on those records, then Alice cannot unequivocally learn Neil's income. 
\end{itemize}

In summary, the training data must be exhaustive for Alice to be sure that at least one of the putative members she has found who share Neil's quasi-identifiers is really Neil. On the other hand, since membership inference to an exhaustive sample is of little value, if Alice turns to attribute inference, she can unequivocally infer a confidential attribute value for Neil only if all putative members sharing Neil's quasi-identifiers share the same (or similar) values for that confidential attribute.  
In looking at the literature on MIAs, most attacks are demonstrated using
training data sets that are not exhaustive and that may contain diverse values for
unknown attributes.

The two conditions have long been studied in the statistical disclosure control (SDC) literature~\cite{sdc-book}: 
\begin{itemize}
    \item The protective effect of non-exhaustive samples is the principle of a well-known SDC method called sampling, in which a sample is released instead of the entire surveyed population. To evaluate the protection provided by sampling, it is relevant to compute the probability that a record is unique in the population ($PU$) given that it is unique
in the sample ($SU$), that is, $\Pr(PU|SU)$. In~\cite{skinner1990disclosure} it was shown that this probability decreases with the sampling fraction, that is, the smaller the sample, the more plausible membership deniability.
\item The protective effect of diversity against confidential attribute disclosure is the principle behind privacy models such as $l$-diversity~\cite{machanavajjhala2007diversity} and $t$-closeness~\cite{li2006t,soria2013differential}: both seek to prevent attribute disclosure
by making sure there is enough diversity of confidential attribute values within each set of records sharing quasi-identifier values. 
\end{itemize}

In fact, it is relatively easy for a model trainer to benefit from the above two protections against MIAs. It is easier to get non-exhaustive than exhaustive training data, and the latter can always be made non-exhaustive by sampling. On the other hand, confidential attributes are naturally diverse and, if there is not enough diversity, it can be enforced by $l$-diversity or $t$-closeness.

Beyond studying the generic limitations of MIAs due to the data used to train the attacked ML models, one can examine the specifics of
the model under attack and the attack method. That is,
what else is needed for an MIA to succeed in the case when the training data happen to satisfy 
exhaustivity and non-diversity. 

In~\cite{jebreel2025critical}, the effectiveness of MIAs
on discriminative machine learning (ML) models is assessed by checking
four requirements: i) the model under attack should not be overfitted
(overfitted models are an easy MIA target, but they do not 
generalize well in their main tasks); ii) the model under attack must have a competitive
test accuracy (attacking an uncompetitive model is not very interesting); iii) the attack must yield reliable membership inference; iv) and the attack
must have a reasonable computational cost. Among the many MIA attacks reviewed by these authors, none can satisfy
these four requirements simultaneously. 

In fact, focusing only on overfitting, ~\cite{dionysiou2023sok} had previously observed
that MIAs on well-generalizable models suffer from
practical limitations that reduce their practicality.
Overall, it would seem that the privacy risks
of machine learning may have been overstated in the literature as far as membership inference attacks are concerned. 

\section{Property inference attacks}
\label{property}

A property inference attack seeks to infer a sensitive {\em global} property
of the data set used to train an ML model, that is, 
a property $P$ of the data set that the model producer did not intend to share. 
This class of attacks was first presented for classifiers
in~\cite{ateniese2015hacking}. They have also been formulated
for deep neural networks in~\cite{ganju2018property}. 

In~\cite{ateniese2015hacking}, a meta-classifier was trained to classify
the target classifier depending on whether it has a certain property $P$
or not. To do this, the attacker trains several {\em shadow classifiers}
on the same task as the target classifier. 
Each classifier is trained on a data set similar to that of the target
classifier, but constructed explicitly to have the property $P$ or not.
Subsequently, the meta-classifier is trained on the sets of parameters
of the shadow classifiers. 

In~\cite{ganju2018property}, it is argued that the above meta-classifier training strategy does not work well
for deep neural networks, due to their complexity and thousands of parameters. The authors explore different feature representations to
reduce the complexity of the meta-classification task. However,
the high-level structure of the attack is the same as
in~\cite{ateniese2015hacking}.
In~\cite{zhou2021property}, a property inference attack against
generative adversarial networks (GANs) is presented. Instead
of training shadow classifiers like in the previous papers, here
shadow GANs are trained. 

From the point of view of privacy, property inference attacks do not entail a
significant risk, because {\em they aim to infer a general property of
the training data set rather than a property specific to a particular target
subject}. That is, {\em property inference attacks are not attribute
inference attacks} trying to infer the value of a confidential attribute
for a target subject. 
By way of illustration, an example of $P$ mentioned in~\cite{ateniese2015hacking}
is whether ``Google traffic was used in the training data'', 
an example mentioned in~\cite{ganju2018property} is whether
``the classifier was trained on images with noise'',
and an example mentioned in~\cite{zhou2021property} is whether
``a GAN is mainly trained with images of white males''.

Even if disclosing such properties was not intended by the model producer
and may cause some embarrassment to them,
the general nature of those properties
can hardly disclose private information on
any of the specific individuals whose data may
have been used for training. The most obvious strategy
to mount an attribute inference attack in machine learning is through
a battery of MIAs each of which hypothesizes a candidate value
for the target subject's confidential attribute 
({\em e.g.}, was the target subject's record with ``Disease=AIDS''
a member of the training data set? was the target subject's record
with ``Disease=Cancer'' a member of the training data set?, and so on).

A scenario where property inference attacks may be more 
privacy-disclosive is federated learning, in case they are used 
to infer properties of the training
data set used by a certain client and those training data refer
to just one or a few subjects. 
Imagine the client is a smartphone and the client's training
data are health measurements on the smartphone owner at different
times; in this specific case, inferring a property of the training data set
can yield a property/attribute of the smartphone owner.

\section{Reconstruction attacks}
\label{reconstruction}

\subsection{Reconstruction attacks previous to ML} 

Dinur and Nissim (DN from now on) developed a formal theory of database reconstruction from a set of query responses in 2003~\cite{dinur2003revealing}.
The authors assume that a database is an $n$-bit
string, that is, it contains records each of which
takes values 0 or 1. They further assume all queries
to be of the form ``How many records in this subset are
0's?'' or ``How many records in this subset are 1's?''. 
In their setting, the response to every query is
computed as the true answer to the query plus an error
$E$ bounded in an interval $[-B,B]$ for some $B>0$. Thus,
the assumption is that query answers are protected
by output perturbation with strictly bounded noise.

According to DN, a database reconstruction is a record-by-record
reconstruction of the original values such that
the distance between the reconstructed values and the original
values is within specific accuracy bounds. DN considered
two types of attackers, one that can ask an exponential
number of queries and one that can only ask a polynomial number of queries, 
and gave results for the reconstructions achievable by
those attackers
as a function of $B$ and the number of queries allowed.
Although such a theoretical framework for database reconstruction
provides very relevant insight, it does not mean that
every database can be uniquely reconstructed. In fact, for a given
set of statistical outputs, there may be several (or even a large
number) of database instantiations compatible with those
outputs~\cite{muralidhar2023database,sanchez2023confidence}.

\subsection{Reconstruction attacks and overfitting in ML}

The problem of reconstructing the data set used to
train an ML model bears some similarities to the database reconstruction
problem just described. 
During machine learning, sometimes the model
memorizes parts of its training data~\cite{feldman2020does}. 
This in turn enables attackers to extract points
from the training data set when given access to the trained model.
Successful reconstruction attacks have been reported for face
recognition models~\cite{fredrikson2015model,zhang2020secret} and neural language
models~\cite{carlini2019secret,carlini2021extracting}.
Although there is no formal framework in the DN style for reconstruction in ML, bounds on the risk of reconstruction have been
proven~\cite{guo2022bounding}.

In fact, (partial) reconstruction of training data
is greatly facilitated if the model is overfitted because, in that case, it memorizes training data.
Beyond being problematic for privacy, 
overfitting is also a great problem for utility,
since overfitted models usually perform poorly
regarding validation (the process of testing how well a trained
model labels new, unseen data).

Regarding potential defenses against
overfitting and, hence, reconstruction, \cite{carlini2019secret} mention that
\begin{quotation}
``such memorization [of training data] is {\em not} due
to overtraining: it occurs early during training, and
persists across different types of models and training
strategies [...] Furthermore, we show that simple,
intuitive regularization approaches such as early-stopping
and dropout are insufficient to prevent unintended memorization.
Only by using differentially-private training techniques,
we are able to eliminate the issue completely, albeit at 
some loss of utility.''
\end{quotation}

Overtraining means training
a model for too many iterations. It may result in overfitting,
which occurs when the model exactly learns the training data set
but is unable to correctly label new, unseen data. 
However, overfitting may also occur in the early stages
of training, that is, without overtraining, such as when
a very large model is trained on a small data set.

In~\cite{blanco2022critical}, it was concluded that
standard anti-overfitting techniques such as regularization
and dropout could outperform DP and achieve a better
utility/privacy/efficiency trade-off in ML training.
The explanation of this seeming contradiction with~\cite{carlini2019secret} 
lies in the details:
\begin{itemize} 
\item \cite{blanco2022critical} tried several combinations
of regularization/dropout and took the one with the best
trade-off between utility, measured as test accuracy, and privacy,
measured as the attacker's (little) advantage in
the standard MIA implementation in TensorFlow Privacy. 
\item In contrast, \cite{carlini2019secret} tried
several anti-overfitting techniques
(regularization, dropout, weight quantization, etc.) but
without attempting to find the best-performing parameterizations.
Also, they measured 
utility as (little) validation loss and privacy
as preventing the recovery of randomly chosen ``canary'' sequences
inserted into the models' training data. 
\end{itemize}

Regarding utility, note that test accuracy and validation loss
are two independent metrics. Whereas the former counts the
number of mistakes/misclassifications, the latter is the distance
between the true labels and the labels predicted by the model.
Low test accuracy means many errors, whereas large validation loss
means large errors.

Regarding privacy, the two above papers and a good deal of the
related literature use MIA-based metrics. 
There are two important factors that influence the success of MIAs:
(i) whether the target
points whose membership is to be inferred are outliers or not and
(ii) how good the MIA techniques employed are. 
Now, the random ``canary'' target sequences
inserted by~\cite{carlini2019secret} in the training data are
likely to be outliers due to their randomness, and hence their membership
may be easy to discover, which gives a pessimistic privacy evaluation.
The TensorFlow Privacy MIA implementation used by \cite{blanco2022critical}
does not rely on the inserting of random
target points into the training data:
it just uses the predictions of the trained model on the target points
to deduce their membership~\cite{boenisch2021attacks}.

\subsection{On the effectiveness of reconstruction in ML}

Using MIAs to assess the effectiveness of reconstruction attacks
may seem reasonable if the training data are tabular.
Let ${\bf D}$ be a training data set with attributes $A_1, A_2, \ldots, A_d$.
Note that in the computer representation of any attribute $A_i$,
the number $|A_i|$ of potential values can be considered finite, even for numerical attributes, due to limited length and precision. Still, $|A_i|$ can be quite large, especially for numerical attributes. 
We can give the following information-theoretic argument to illustrate
the complexity of exhaustively trying all possible values. 
Assume that the information content
of an item $X$ (record in the case of tabular data,
but also unstructured text, image, etc., for non-structured data) one
wishes to reconstruct is $H(X)$ bits, where $H$ is Shannon's entropy. 
Then discovering $X$ by exhaustive search is equivalent to discovering a random cryptographic key
of $H(X)$ bits. If $H(X)$ is, say, 64 or more bits, this is known to be computationally infeasible.

This gives two scenarios:
\begin{enumerate}
\item {\em Total reconstruction}.
Assume that the attacker has unlimited resources or, better, 
that the number of potential
values $|A_i|$ of every attribute $A_i$ is relatively small.
In this case, the attacker could mount
an MIA for each possible combination of attribute values, to check
whether that combination was part of ${\bf D}$.
After $\prod_{i=1}^d |A_i|$ MIAs have been performed and if
they are effective, the attacker has reconstructed
the entire training data set $D$.
\item {\em Partial reconstruction}.
If the attacker's resources are insufficient
to pursue total reconstruction, then they can select
a subset of possible combinations of attribute values
and mount MIAs only for those combinations. 
This can be viewed as a {\em guessing exercise}
that may lead to a partial reconstruction of ${\bf D}$ (if the guesses, that is,
the candidate combinations
of attribute values, 
are classified as members and are really members of ${\bf D}$). 
The attacker would favor those combinations
deemed to be the most likely from the semantics
of attributes, {\em e.g.} if there is an attribute {\em Age} 
and an attribute {\em Job}, the only plausible combination
of {\em Age}=10 is with {\em Job}=`student'. Note that
betting on the most common combinations gives less interesting
reconstruction results for the attacker: outlier combinations
are more privacy-sensitive and thus interesting to the attacker
than very common combinations. 
\end{enumerate}

It must be taken into account that state-of-the-art MIAs
offering the best membership detection, such as LiRA~\cite{carlini2022membership},
require training several shadow models to estimate the
distribution $\Delta_{in}$
of models trained on data sets containing the target point
and the distribution $\Delta_{out}$ 
of models trained on data sets {\em not} containing
the target point. 
Thus, each MIA incurs a substantial computation cost. 

Furthermore, especially in generative ML, 
training data are often non-tabular. For example, they 
are unstructured text or multimedia. 
Clearly, for non-tabular training data such as images
or unstructured text, mounting an MIA
to test whether each potential image or each potential unstructured
text was part of the training data set
${\bf D}$ seems quite unreasonable. 
In the case of generative AI, one can resort to prompting
for certain personal data or copyrighted content
rather than mounting MIAs, in order to
find out whether the model saw those items at training time.
But in fact, this prompting amounts to a guessing exercise like those described
above under partial reconstruction. 
In fact, the empirical study~\cite{duan2024membership} shows that
MIAs on pre-trained LLMs are barely better than random guessing, even
though fine-tuned LLMs are far more vulnerable to MIAs. That is, MIAs 
are more effective at inferring membership on the data used for fine-tuning than on the data used for pre-training.
Regarding cost, although guess prompting
is almost free on the user's side, the computational
cost is high in terms of LLM inference on the LLM manager's side.

In~\cite{liu2025data} a systematic evaluation of data reconstruction
attacks and defenses is presented, where the reconstruction
attacks considered are no longer MIAs, but {\em gradient inversion
attacks}. 
Gradient inversion attacks~\cite{zhang2022survey} 
attempt to recover training points from gradients. 
They are mostly designed for federated learning (FL),
because they require
knowledge of the gradients computed during training.
In fact, in FL, the server receives the gradients
from the clients and can mount a gradient inversion
attack and try to reconstruct the local training
data for one or more clients~\cite{shi2024dealing}.
If all clients receive all gradients, then clients can also
behave maliciously and mount a gradient inversion attack
to reconstruct the local data of a certain target client.
The study~\cite{ovi2023comprehensive} reviewed
gradient inversion attacks
against FL, as well as potential defenses based on
mixed precision and quantization, gradient pruning,
and differential privacy. 
They concluded that some of these defenses are effective
and involve only slight accuracy drops.
In centralized learning, where the attacker only sees
the trained model, gradient inversion attacks are not applicable.

If reconstruction based on MIAs is problematic for the reasons
above, reconstruction without MIAs suffers from a major weakness:
{\em there is no numerical decision criterion in a realistic
case in which the attacker has no access to the
actual training data}. In other words, whereas
in an MIA there is some kind of threshold that allows deciding
whether a target point is a member or a non-member
(although this decision may be in error), in a reconstruction
attack there is no objective criterion to decide whether
the putative reconstructed data belong to the training data set.
For example, the fact that a gradient inversion attack produces
a meaningful image does not necessarily mean that this image
was part of the training data. Also, what ``meaningful'' means
is debatable. One could certainly use an MIA to decide whether
the putative reconstructed data were really in the training data,
but this has the drawbacks of MIAs enumerated in Section~\ref{MIA}.

Admittedly, there are situations in which it may be easier to
make a decision on putative reconstructed data. This
is the case for reconstruction attacks on machine
unlearning.
In unlearning, a trained model is updated to cause it to
``forget'' one or more data points, {\em e.g.}, to implement
the right to be forgotten enshrined in the GDPR, or because
those data points are subject to copyright.
In~\cite{bertran2024reconstruction}, 
a reconstruction attack is described for the case in which the trained
model is a simple one. The attack exploits the model updates
to estimate the unlearned data point.
However, even if the attack is quite successful
according to the experiments reported 
in~\cite{bertran2024reconstruction}, success is determined by comparing
against the ground truth of the unlearned data point,
which would not be available to an attacker in a real world situation. 
Possible defenses are discussed in~\cite{domingo2025defenses}.

\section{Conclusions}
\label{conclusion}

Our analysis casts doubts on the effectiveness of privacy attacks against ML
in real-world conditions:
\begin{itemize}
\item MIAs suffer from limitations due to the data the target models have been trained on (non-exhaustivity, diversity of confidential attribute values). In addition, they may also suffer limitations that arise from the nature of the attacked models and the attack methods.
\item Property inference attacks aim to infer a general property
of the training data set, rather than a property specific to a particular
data subject. For that reason, they do not achieve
attribute disclosure for any particular subject and hence do not
pose substantial privacy
risks to subjects, 
except in specific federated learning scenarios where all of 
a client's training data refer to one or a few subjects. 
These attacks are more relevant to audit the potential biases
or insufficiencies of the training data used by the model producer.
\item Reconstruction attacks based on MIAs have a very significant
cost, as they involve mounting an MIA for each data point whose membership
in the training data is to be decided. 
Thus, they are only practical for tabular training data where attributes
have a limited range of potential values, and even in that case they are 
more suited for partial than total reconstruction.
Besides, MIA-based reconstruction is
also subject to the shortcomings identified for MIAs themselves.  
\item Reconstruction attacks based on gradient inversion are
those that are used when training data are multimedia or unstructured
text, as is the usual case in generative ML. However, such attacks are
applicable only when the attacker has access to the gradients computed by
the victim during the learning process. In practice, this restricts
the applicability of these attacks to federated or otherwise decentralized
learning. Furthermore, deciding whether a putative reconstructed
data point was really a member of the training data is difficult
if the attacker does not have access to the original training data
(which is the usual case in the real world). Certainly, MIAs can be used to make this membership decision, but this inherits
the shortcomings of MIAs described above.
\end{itemize}

All in all, the current
real-world privacy risks incurred by machine learning
seem less serious than what is usually assumed in the literature.
Therefore, privacy defenses that entail severe utility loss, such
as differential privacy, may be often unnecessary. The good side
of all this is that trustworthy machine learning may be easier
to implement than assumed so far, at least with respect to privacy. 
This is good news for jurisdictions like the European
Union that struggle to reconcile
strong AI regulations with the competitiveness of their AI industry.  

\section*{Acknowledgments}

This work was partly funded by the Centre International de Math\'ematiques et d'Informatique de Toulouse
(CIMI), the
Government of Catalonia (ICREA Acad\`emia Prize to J. Domingo-Ferrer), 
MCIN/AEI/ 10.13039/501100011033 and ``ERDF A way of making Europe'' under grant PID2021-123637NB-I00 ``CURLING'', and INCIBE and European Union NextGenerationEU/PRTR (project ``HERMES'' and INCIBE-URV Cybersecurity Chair).

\bibliographystyle{plain}

\end{document}